# Quantum theory of securities price formation in financial markets

(with application to spread and return probability distribution modeling)

Jack Sarkissian

Managing Member, Algostox Trading

email: jack@algostox.com

Abstract

We develop a theory of securities price formation and dynamics based on quantum approach without presuming any similarities with quantum mechanics. Disorder introduced by trading environment leads to probability distribution of returns that is not a smooth curve, but a speckle-pattern fluctuating in both price coordinate and time. This means that any given return can at times acquire a substantial probability of occurring while remaining low on average in time. Still, due to local character of order interaction during price formation the distribution width grows smoothly, has a minimum value at small time scale and behaves $\sim \sqrt{t}$ at large time scale. Examples of calibration to market data, both intraday and daily, are provided.

## 1. Introduction

It is commonly accepted that security prices follow a diffusion process. That process can be free diffusion, mean-reverting, with or without memory, but it's always some form of diffusion. Accordingly, probability distribution of returns obeys the diffusion equation, and for free diffusion its width scales as $\sqrt{t}$ with time. This theory works well for a large time scale and on liquid securities. Many financial institutions work within this framework, applying it to making trade decisions, derivatives pricing, SLB[1], and risk management.

Picture is different at time scale that is as small as the typical time between transactions. Price uncertainty stops scaling down and does not approach zero anymore. It stays at the value relevant enough for the security [1-3] and cannot be reduced further. That value is represented by bid-ask spread of the order book. Additionally, the return distribution substantially deviates from Gaussian and becomes more fat-tailed [4]. Clearly the diffusion theory breaks, and a new theory is needed that can relate to spread as its intrinsic characteristic. That theory must have the diffusion theory as a limiting case for large time scale.

Let us specify some notions. In financial markets buying and selling orders are lined up on an exchange, and the record of current orders is called an order book. An example of order book is shown in Fig. (1). The trading parties wait in line for a matching order, and when such order arrives a transaction is recorded. Until a transaction occurs there is no single price to the security. Instead, there is a spectrum of prices, that potentially represents the price. Spread is the difference between the lowest selling and highest buying prices:

$$\Delta = s_{ask} - s_{bid}, \qquad (1)$$

where $s_{ask}$ is the best "ask" and $s_{bid}$ is the best "bid". By its nature, spread represents the risk premium that is supposed to cover price uncertainty during the liquidation period.

---
[1] Securities lending and borrowing

| Bid | | Ask | |
|---|---|---|---|
| Price | Size | Price | Size |
| 27.83 | 100 | 27.87 | 100 |
| 27.82 | 100 | 27.9 | 100 |
| 27.8 | 200 | 27.95 | 1000 |
| 27.79 | 600 | 28.15 | 300 |
| 27.78 | 100 | 28.2 | 400 |

Fig. 1. Sample order book.

Spread has been a subject of active research in recent years. The most common approach is to obtain it as a result of modeling processes in the order book. Once order arrival, cancellation and execution are adequately modeled, spread is obtained by direct calculation from Eq. (1) [5,6]. Another approach is to obtain spread as the optimal value from market maker's perspective [7,8]. This approach allows to calculate spread from microstructural parameters, but also depends on the degree of market maker's risk aversion.

We believe that theory of spread and price formation should begin with understanding the nature of price. Security price as a specific number is not an object that exists continuously and independently. It exists only when it is measured: at the time of transaction. Without a transaction, one can just say that the price is locked between best bid and best ask prices. However, sometimes one of these prices may not exist. At times both prices may not exist. We see that on microstructural level we can at best talk about price localization in a certain region. Price uncertainty disappears only at the time of transaction. We can say that **every financial transaction is an elementary act of price measurement**. Measurement leads to selection out of the entire price spectrum according to probability of each price and probability of the security to be found in the state with that price [9]. These considerations point to quantum nature of price formation.

Application of quantum mechanics to finance is not surprising: quantum mechanics is probabilistic, and so are financial markets. A number of attempts has been made to describe price behavior with methods of quantum mechanics [10-19]. Some of them map the known quantum mechanical solutions to finance, thus relying on a postulate that quantum mechanics applies to finance unchanged. Others attempt to rederive quantum equations as they apply to finance. Although these models yield results with characteristics of financial markets, we feel that more evidence is needed for how well they describe the observed market data and how well they can be calibrated and used in an institutional setting.

At large time scale $\Delta t \gg \tau_0$, $\tau_0$ being the average time between transactions, price variations are usually much larger than the spread, and price spectrum as well as the trading process can be considered continuous. At such time scale a valid theory of price formation must go back to stochastic framework and result in $\sim\sqrt{t}$ widening of probability distribution provided by the diffusion equation. Motivated by this argument, some researchers see the solution in attempting to extract Schrodinger equation from the diffusion equation [18]. We believe that stochastic framework should be obtained as a limiting case from quantum framework, and not the other way around.

In this paper we further develop quantum framework originally proposed by us in [1]. In that work we showed that the two-level quantum coupled-mode model adequately describes the statistics of spread and mid-price. Further development of the model in [2] showed that the model is capable of explaining the relationship between spread, trading volume and volatility. These results make this theory very promising. Here we do not pursue strict mathematical completeness, rather we are concerned with building analytical base corresponding to the nature of processes and useful in practical applications.

We will refer to "spread" as a broader concept that can be defined in different ways for different levels of data specification. For example, if the data represents an instantaneous snapshot of order book then the spread is literally the difference between the best offer and the best bid in the book. We can also consider the spread to be the difference between the effective offer and effective bid prices, which are represented by order weighted averages of top $N$ levels of order book. Lastly, if the data is represented by OHLC[2] time series, one can consider the spread to be the difference between high and low prices. These notions are formulated more accurately in [1].

For simplicity, but without loss of generality, we will discuss only equities here (stocks). Similar considerations are applicable to other asset classes, such as fixed income, currency, commodities, and futures.

## 2. Price operator and evolution equation

*Price operator and its properties*

The basic element of our approach is the probability amplitude $\psi$, which describes the state of the security, and whose absolute value squared represents the probability of finding the security in the given state:

$$p = |\psi|^2 \qquad (2)$$

Security prices are governed by the *price operator* $\hat{S}$. Price operator's eigenfunctions represent the pure states in which the security can be found, and eigenvalues represent the spectrum of prices that the security can attain in each corresponding state with quantum number $n$:

$$\hat{S}\psi_n = s_n \psi_n \qquad (3)$$

To guarantee that prices are real numbers the price operator $\hat{S}$ must be Hermitian. Price spectrum is determined by the following equation

$$\det\|s_{mn} - s_n \delta_{mn}\| = 0 \qquad (4)$$

---

[2] OHLC are the traditional open-high-low-close bars

Matrix elements of the price operator fluctuate in time, which introduces random properties to eigenvalues and eigenfunctions:

$$\hat{S}(t + \delta t) = \hat{S}(t) + \delta \hat{S}(t) \tag{5}$$

and therefore:

$$s_n(t + \delta t) = s_n(t) + \delta s_n(t) \tag{6}$$

For a two-level system Eq. (3) takes the following form:

$$\psi = \begin{pmatrix} \psi_{ask} \\ \psi_{bid} \end{pmatrix} \quad \text{and} \quad \begin{pmatrix} s_{11} & s_{12} \\ s_{12}^* & s_{22} \end{pmatrix} \begin{pmatrix} \psi_{ask} \\ \psi_{bid} \end{pmatrix} = s_{ask/bid} \begin{pmatrix} \psi_{ask} \\ \psi_{bid} \end{pmatrix} \tag{7}$$

Given this, the bid and ask prices can be obtained as

$$s_{ask} = s_{mid} + \frac{\Delta}{2} \quad \text{and} \quad s_{bid} = s_{mid} - \frac{\Delta}{2} \tag{8a}$$

$$s_{mid} = \frac{s_{11} + s_{22}}{2} = \frac{s_{bid} + s_{ask}}{2} \quad (\text{mid} - \text{price}) \tag{8b}$$

$$\Delta = \sqrt{(s_{11} - s_{22})^2 + 4|s_{12}|^2} \quad (\text{spread}) \tag{8c}$$

The fluctuating matrix elements $s_{ik}$ of price operator can be parametrized as follows:

$$s_{11}(t + dt) = s_{mid}(t) + \sigma dz + \frac{\xi}{2} \tag{9a}$$

$$s_{22}(t + dt) = s_{mid}(t) + \sigma dz - \frac{\xi}{2} \tag{9b}$$

$$s_{12}(t + dt) = \frac{\kappa}{2} \tag{9c}$$

where $dz \sim N(0,1)$, $\xi \sim N(\xi_0, \xi_1)$ and $\kappa \sim N(\kappa_0, \kappa_1)$. In such setup the mid-price and the spread are given by equations:

$$s_{mid}(t+dt) = s_{mid}(t) + \sigma dz \tag{10a}$$

$$\Delta = \sqrt{\xi^2 + \kappa^2} \tag{10b}$$

As we can see, the mid-price simply follows a Gaussian process with volatility $\sigma$. Spread, on the other hand, behaves as quantum-chaotic quantity [1, 20], whose statistics matches the observed spread statistics quite well both on bid-ask micro-level and on bar data level, see Figs. 2a, 2b.

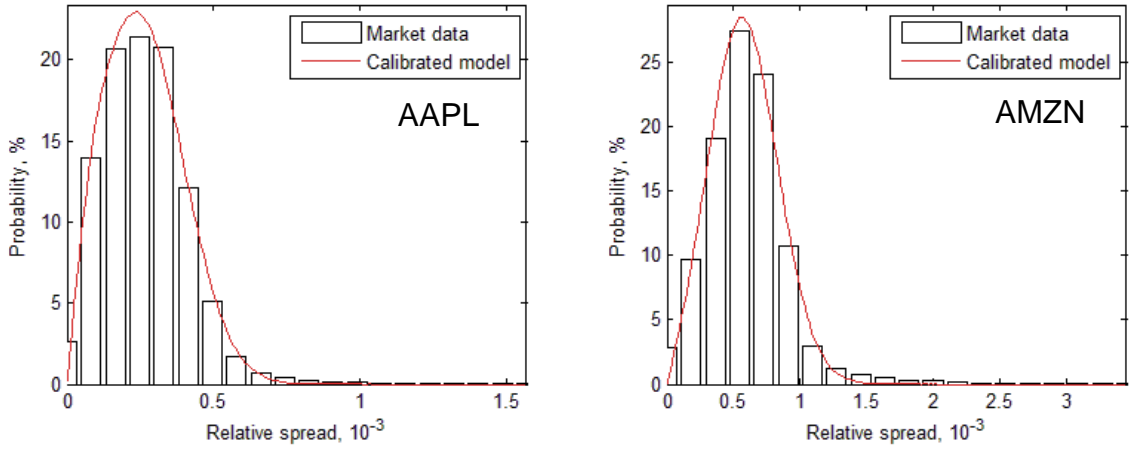

Fig. 2a. Calibration of Coupled-Wave Model to bid-ask data for AAPL and AMZN, [1]

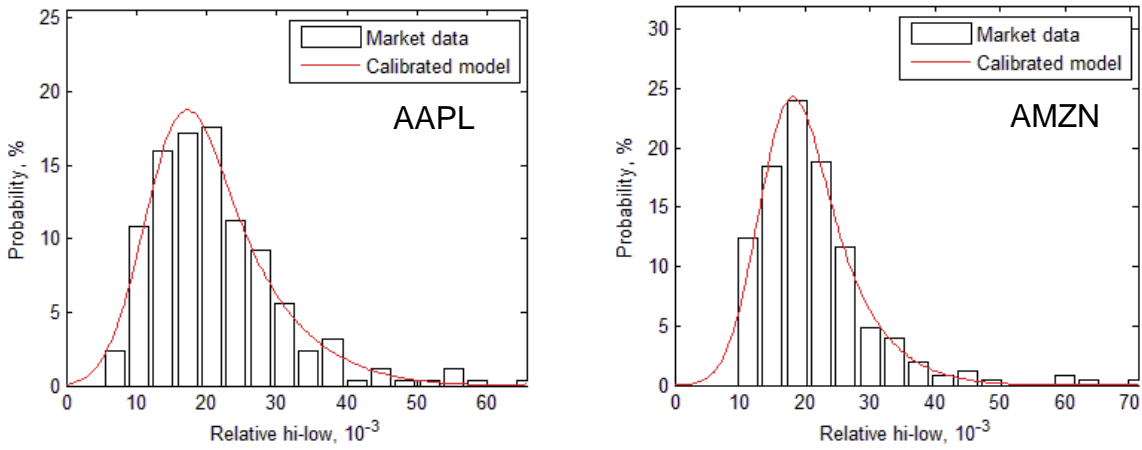

Fig 2b. Calibration of Coupled-Wave Model to bar data for AAPL and AMZN, [1]

*Price dynamics*

Evolution equation for the probability amplitude can be obtained from the following consideration. Total probability is conserved and equal to one at all times:

$$\sum_n |\psi_n|^2 = 1 \tag{11}$$

Then, differentiating this in $t$:

$$\sum_n \left( \psi_n^* \frac{\partial \psi_n}{\partial t} + \psi_n \frac{\partial \psi_n^*}{\partial t} \right) = 0 \tag{12}$$

The most general linear equation satisfying this condition is[3]:

$$i\tau s \frac{\partial \psi}{\partial t} = \hat{S} \psi \tag{13}$$

where we factored out time constant $\tau$ and price constant $s$ to be able to identify $\hat{S}$ with the price operator. If matrix elements of $\hat{S}$ remain constant over time step $\Delta t$, we can write the solution of Eq. (13) as

$$\psi(\Delta t) = e^{-i \frac{\hat{S}}{s} \frac{\Delta t}{\tau}} \psi(0) \tag{14}$$

For a timeframe consisting of $N$ time steps we can write:

$$\psi^{(r)}(t) = e^{-i \frac{\hat{S}_N^{(r)}}{s} \frac{\Delta t_N}{\tau}} e^{-i \frac{\hat{S}_{N-1}^{(r)}}{s} \frac{\Delta t_{N-1}}{\tau}} \dots e^{-i \frac{\hat{S}_1^{(r)}}{s} \frac{\Delta t_1}{\tau}} \psi(0) \tag{15}$$

---

[3] Stricter derivation of this equation is given in [21]

where $\hat{S}^{(r)}$ is a particular realization of matrix $\hat{S}$ in time and $\psi^{(r)}(t)$ means a $\psi(t)$ resulting from that realization. To find the expected probability distribution one has to average over the ensemble of all realizations:

$$P(t) = \langle |\psi^{(r)}(t)|^2 \rangle_r \tag{16}$$

Elementary steps in Eq. (15) can be performed analytically using the Lagrange–Sylvester formula for matrix exponential, so that:

$$\psi(\Delta t) = \left[ \sum_{j=1}^{n} e^{\frac{s_j}{s} \frac{\Delta t}{\tau}} \prod_{\substack{k=1 \\ k \neq j}}^{n} \frac{\hat{S} - s_k \hat{I}}{s_j - s_k} \right] \psi(0) \tag{17}$$

where $\hat{I}$ is the identity operator and price spectrum $\{s_j\}$ is determined by equation Eq. (4).

Let us say a few words about price spectrum. If the security could only attain a single price value, then all diagonal elements of price operator would be equal to the security price and all off-diagonal elements would be equal to zero:

$$s_{mn} = s\, \delta_{mn} \tag{18}$$

Off-diagonal elements are responsible for interaction between price levels. Since distant quotes generally do not affect pricing, a nearest neighbor approximation can be used for many applications:

$$s_{nn} = s + \xi_n, \qquad \xi_n \sim N(0, \xi) \tag{19}$$

$$s_{n+1,n} \text{ and } s_{n-1,n} \sim N(0, \kappa) \tag{20}$$

Calibrated with such model 1-minute and 1-day return spectrums are shown for LULU and AAPL tickers in Fig. 3.

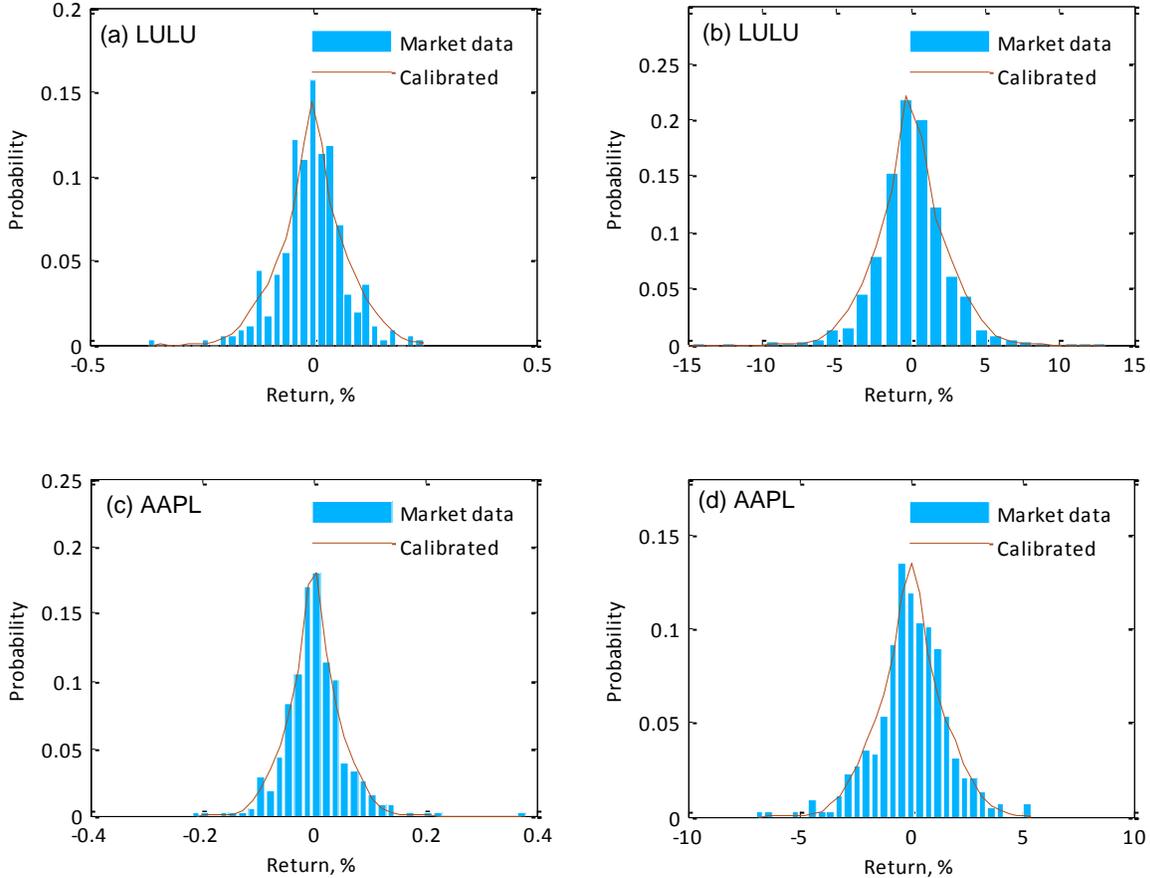

Fig. 3. Return spectrum and probability distribution for LULU and AAPL: (a) 1-minute returns for LULU and (b) daily returns for LULU, (c) 1-minute returns for AAPL and (d) daily returns for AAPL. Intraday returns exhibit stronger non-Gaussian features that are captured by the model.

Main qualitative features of the dynamics of probability distribution resulting from Eq. (15) can be observed in Figs. 4-7. We see form Fig. 4 and Fig. 5 that probability distribution is no longer represented by a smooth curve. Though localized in price, it is erratic and its shape is represented by a speckle-pattern [22]. Dynamics of probability distribution is erratic too, though its localization width grows gradually. Fig. 5b shows evolution of probabilities of various price returns: 0% (unchanged), –10%, and –20%. We see that any return – big and small – can at times have a good chance of occurring. Probability of large price shifts can sometimes build up to substantial values, making space for the so called "Black Swans". Risk measures, shown in Fig. 5c, exhibit a $\sqrt{t}$-like behavior. They are also erratic and can deviate from Gaussian values, which is evidenced by risk ratios on Fig. 5d (1.64 for 95% VaR and 2.06 for 95% Expected Shortfall). This deviation is especially large at smaller time scale and diminishes with time. Ensemble average produces smooth curves, shown in Figs. 6,7. Even though an ensemble allows to take into account all possibilities, in reality we must face only a single realization, with its erratic behavior.

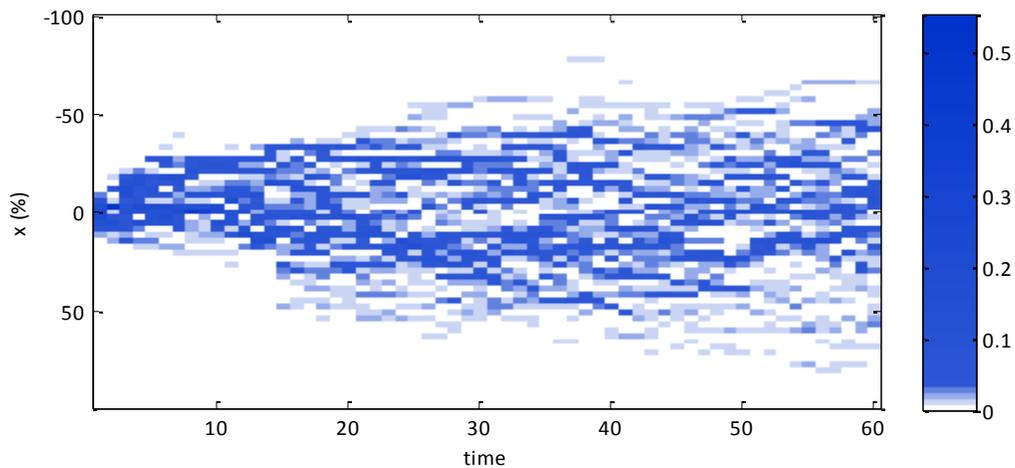

Fig. 4. Dynamics of probability distribution in a single realization

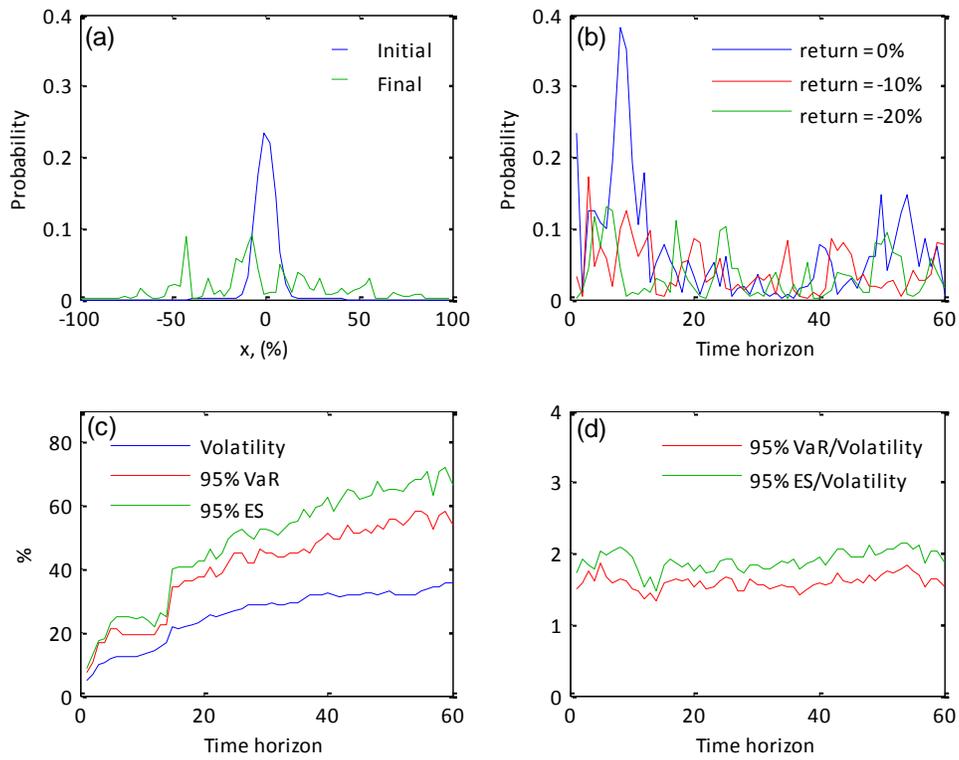

Fig. 5. Dynamics of probability distribution in single realization: (a) initial and final probability distributions, (b) dynamics of probabilities, (c) risk measures, and (d) risk ratios

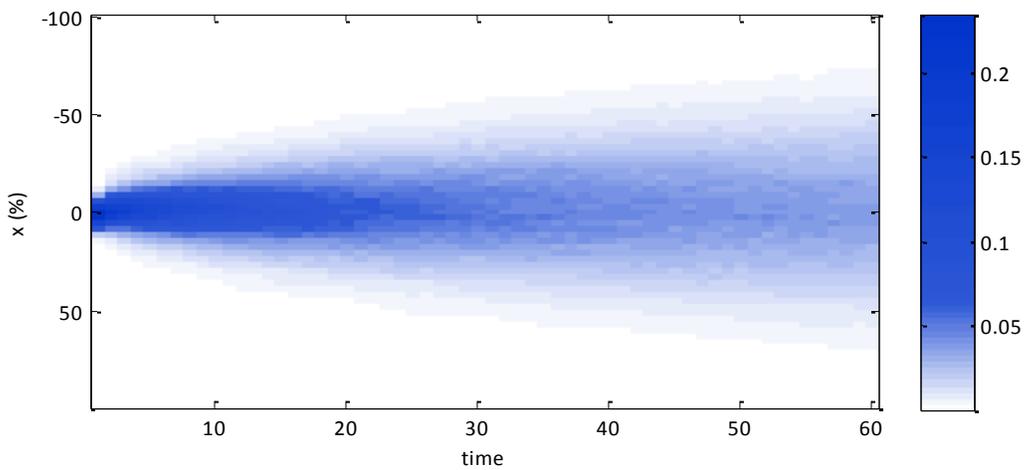

Fig. 6. Ensemble average dynamics of probability distribution.

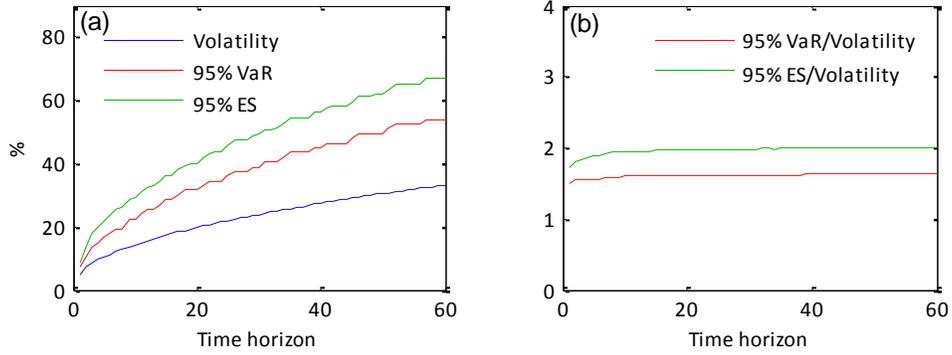

Fig. 7. Ensemble average dynamics of probability distribution:

(a) risk measures and (b) risk ratios

*Evolution equation in coordinate representation*

Let us now write the evolution equation in continuous coordinate space. We will use the logarithmic price $x = ln(s)$ for coordinate, so that $\psi_n$ represents the probability amplitude for finding the security in a state with $x_n = \ln(s_n)$. We will begin by transforming Eq. (3) into a master-equation [23]. This equation in matrix form reads:

$$i\tau s \frac{d\psi_n}{dt} = \sum_m s_{nm} \psi_m \tag{21}$$

Let us now construct a matrix $\hat{Q}$ in the following way:

$$q_{nm} = \frac{s_{mn}}{s} \quad \text{for } n \neq m \tag{22a}$$

$$q_{nn} = \frac{\sum_l s_{ln}}{s} \tag{22b}$$

Then matrix elements of $\hat{S}$ and $\hat{Q}$ are connected through the following relation:

$$s_{nm} = s\left(q_{mn} - \delta_{nm}\sum_{l \neq n} q_{nl}\right) \tag{23}$$

where $s$ is factored out so that matrix coefficients do not scale with price. Eq. (21) now transforms into:

$$i\tau \frac{d\psi_n}{dt} = \sum_m q_{mn}\psi_m - \sum_{m \neq n} q_{nm}\psi_n \tag{24}$$

This equation has a form of a master-equation. In nearest-neighbor approximation we have:

$$i\tau \frac{df_n}{dt} = q_{n-1,n}f_{n-1} - (q_{n,n-1} + q_{n,n+1})f_n + q_{n+1,n}f_{n+1} \tag{25}$$

where we used the $\psi_n = e^{-iq_{nn}\frac{t}{\tau}}f_n$ substitution. Let us introduce the functions $\mu$ and $\gamma$, such that

$$i\mu_n = \Delta x \left(q_{n,n-1} - q_{n,n+1}\right) \tag{26a}$$

$$\gamma_n = \frac{\Delta x^2}{2}\left(q_{n,n-1} + q_{n,n+1}\right) \tag{26b}$$

Elements $q_{nm}$ can be expressed through these functions as

$$q_{nm} = \kappa_n \delta_{n-1,m} + \kappa_m^* \delta_{n+1,m} \tag{27a}$$

$$\kappa_n = \frac{\gamma_n}{\Delta x^2} + i\frac{\mu_n}{2\Delta x} \tag{27b}$$

Performing transition to continuous form we arrive at the following Fokker-Planck equation:

$$i\tau \frac{\partial f}{\partial t} = i \frac{\partial (\mu f)}{\partial x} + \frac{\partial^2 (\gamma f)}{\partial x^2} \tag{28}$$

Here $\mu = \mu(x,t)$ and $\gamma = \gamma(x,t)$ are generally functions of coordinate and time. If $\mu$ and $\gamma$ fluctuate in time, then an ensemble average should be taken according to Eq. (16) in order to obtain a complete probability distribution. The $\frac{\partial}{\partial x}$ term is responsible for the drift of probability amplitude in coordinate space, and the $\frac{\partial^2}{\partial x^2}$ term is responsible for dispersion. We do not make a statement that $\gamma$ necessarily has to have a definite sign, since we are developing a theory without drawing any prior similarities to traditional quantum mechanics. Properties of $\mu$ and $\gamma$ have to be established by further research as it relates to financial markets.

Equation Eq. (28) combined with initial condition $f(x, t=0) = f_0(x)$ formulates an initial value problem that describes evolution of probability amplitude of security's logarithmic price (or effectively returns) with time.

## 3. Probability evolution

Let us now apply Eq. (28) to study evolution of price probabilities. We will assume the initial price to be localized with width $w_0$ according to Gaussian probability distribution, so the probability amplitude is:

$$f(x, 0) = \frac{1}{\sqrt[4]{2\pi w_0^2}} e^{-\frac{x^2}{4w_0^2}} \tag{29}$$

That way $\int_{-\infty}^{\infty} p(x) dx = \int_{-\infty}^{\infty} |f(x,0)|^2 dx = 1$. We will model the fluctuating parameters $\mu(x,t)$ and $\gamma(x,t)$ piecewise in time, refreshing them after each time step:

$$\mu(x,t) = \mu_k(x), \qquad \text{for } t_{k-1} \le t < t_k \tag{30}$$

$$\gamma(x,t) = \gamma_k(x), \qquad \text{for } t_{k-1} \le t < t_k \tag{31}$$

To show the effect of different terms of Eq. (28) on probability evolution let us first consider a simple, but not so applicable case. If price remains localized in a narrow price interval, spatial inhomogeneity of $\mu$ and $\gamma$ will not affect it much. In other words, if $\frac{1}{\mu}\frac{\partial \mu}{\partial x}, \frac{1}{\gamma}\frac{\partial \gamma}{\partial x} \ll \frac{1}{w}$ and $\frac{1}{\gamma}\frac{\partial^2 \gamma}{\partial x^2} \ll \frac{1}{w^2}$, we can assume that parameters $\mu$ and $\gamma$ are independent of price. Furthermore, we will assume that $\gamma$ is constant in time: $\mu = \mu(t)$ and $\gamma = \gamma_0$. These assumptions reduce Eq. (28) to

$$i\tau \frac{\partial f}{\partial t} = i\mu(t)\frac{\partial f}{\partial x} + \gamma_0 \frac{\partial^2 f}{\partial x^2}, \tag{32}$$

where $\mu(t)$ fluctuates in time with standard deviation $\sigma_\mu$: $\mu(t) \sim N(0, \sigma_\mu)$. Solving Eq. (32) for one realization over the interval $\Delta t$, we get:

$$f^{(r)}(x, \Delta t) = \frac{1}{\sqrt[4]{2\pi w^2}} e^{-\frac{\left(x + \frac{\mu^{(r)} \Delta t}{\tau}\right)^2}{4w^2} + i\phi} \tag{33}$$

where $\phi$ is some phase, unimportant here, and width is

$$w = w_0 \sqrt{1 + \left(\frac{\gamma}{w_0^2}\right)^2 \left(\frac{\Delta t}{\tau}\right)^2} \tag{34}$$

Finally, taking an ensemble average of Eq. (33) over all possible realizations of $\mu$, we obtain the final average probability distribution:

$$P(x) = E_r\left[|f^{(r)}(x,t)|^2\right] = \frac{1}{\sqrt{2\pi w(t)^2}} e^{-\frac{x^2}{2w(t)^2}} \tag{35}$$

where width is given by:

$$w(t) = w_0 \sqrt{1 + \frac{\sigma_\mu^2}{w_0^2}\frac{t}{\tau} + \left(\frac{\gamma}{w_0^2}\right)^2 \left(\frac{t}{\tau}\right)^2} \tag{36}$$

Thus, when inhomogeneity is small probability distribution maintains Gaussian shape during evolution, both in a single realization and in ensemble average. An example of this can be seen in Figs. 8,9, that show that risk ratios constantly remain on their Gaussian levels: 1.64 and 2.06.

In this model temporal behavior of the width changes from square root behavior linear, which is not physical. This is not surprising, since we made assumptions that make the solution valid only for small timeframe, in which inhomogeneity of trading environment does not come into play. More specifically, equation Eq. (36) can only be applied at timescale $t \lesssim \left(\frac{w_0 \sigma_\mu}{\gamma}\right)^2 \tau$.

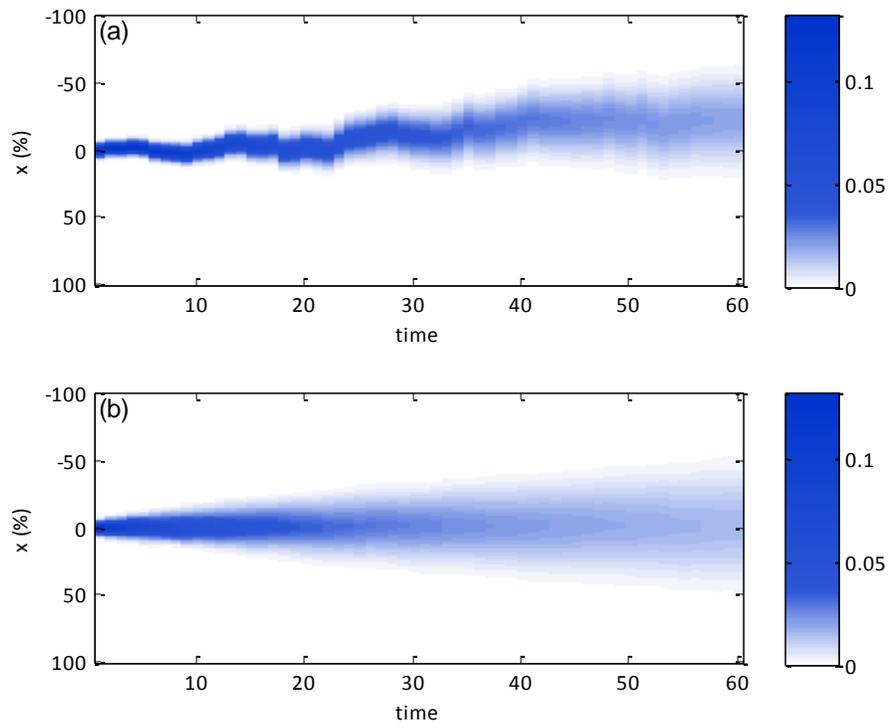

Fig. 8. Dynamics of probability distribution:

(a) single realization (b) ensemble average over 100 realizations

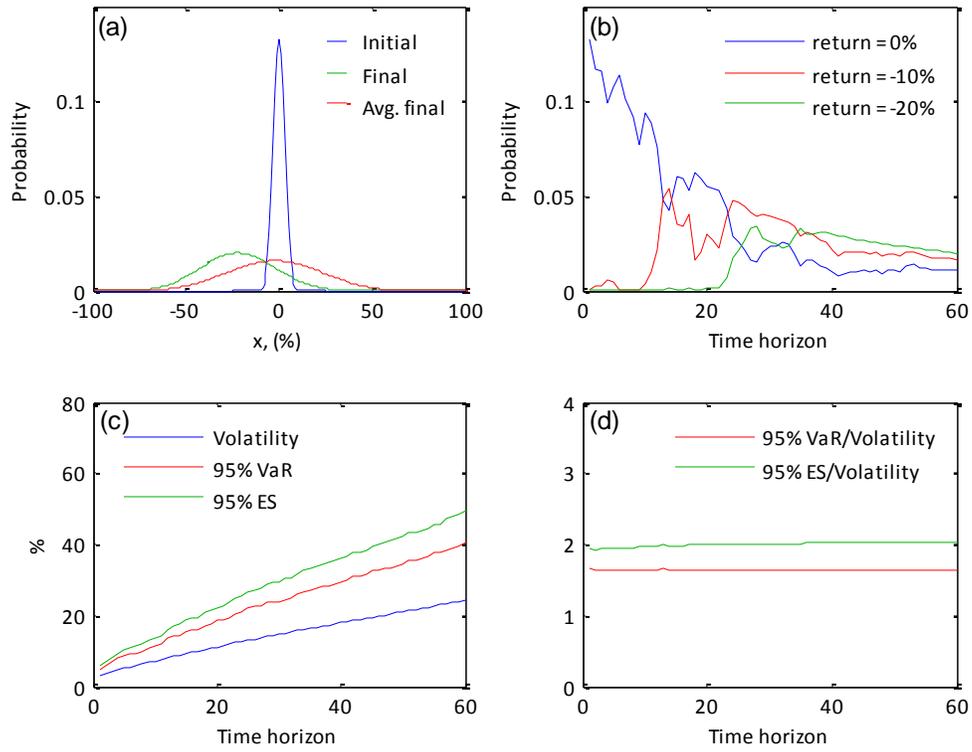

Fig. 9. Dynamics of probability distribution:

(a) initial, final, and average final probability distributions, (b) dynamics of probabilities, (c) risk measures, and (d) risk ratios

*Probability evolution in random trading environment*

While Eq. (28) allows to study numerous models with various types of dependencies of $\mu$ and $\gamma$, and having various log-price structures, let us consider a model with large disorder, leading to complete randomization of phase at each step of evolution. Let $\mu$ fluctuate in time: $\mu = \mu(t)$, and $\gamma$ fluctuate in both time and price: $\gamma = \gamma(x, t)$, and let $\gamma$ have granularity size $\epsilon = \Delta x$ in price, so that values in the neighboring price levels are uncorrelated. We thus have for Eq. (27b): $\mu_n = \mu \sim N(0, \sigma_\mu)$ and $\gamma_n = N(0, \sigma_\gamma)$. Standard deviations $\sigma_\mu$ and $\sigma_\gamma$ must be chosen so that the eigenvalues of the price operator match the 1-step price spectrum.

Evolution of probability amplitude in such environment can be approached by representing it as a superposition of probability amplitudes at different price levels:

$$\psi(x,t) = \int_{-\infty}^{\infty} a_\rho(x')\psi_\epsilon(x' - x, t)dx' \qquad (37)$$

where $\psi_\epsilon$ is localized within the range $\epsilon$, and $a_\rho$ is an envelope function with width $\rho$, such that relation $w^2 = \rho^2 + \epsilon^2$ is maintained.

Width of propagated amplitude $\psi_\epsilon(x, t + \Delta t)$ is described by Eq. (36) and grows if $\tau$ decreases. However, the maximum value it can take in a single step is $\sim 3\epsilon$, because only the neighboring levels interact. At small enough $\tau$, such that $\tau \gtrsim \frac{\gamma \Delta t}{\pi w^2}$ the level of disorder is so high that phase begins to change randomly with price and its values cover the entire interval $(-\pi; \pi]$. At this point we can neglect interference and present the propagated distribution as:

$$|\psi(x, t + \Delta t)|^2 = \int_{-\infty}^{\infty} a_\rho^2(x')|\psi_\epsilon(x' - x, t + \Delta t)|^2 dx' \qquad (38)$$

As a result, the probability distribution only widens randomly by about one level in each evolution step:

$$w^2(t + \Delta t) = w^2(t) + (\beta \epsilon)^2 \qquad (39)$$

where $\beta \sim 1$. Thus, when $\tau$ becomes too small width $w$ levels off and ceases its direct dependence on $\sigma_\mu$ and $\sigma_\gamma$. No matter how much further $\tau$ is decreased, it will no effect on dynamics of $\psi$. We can see that effect in Fig. 10, presenting the $w(T)$ as it changes with $\frac{1}{\tau}$.

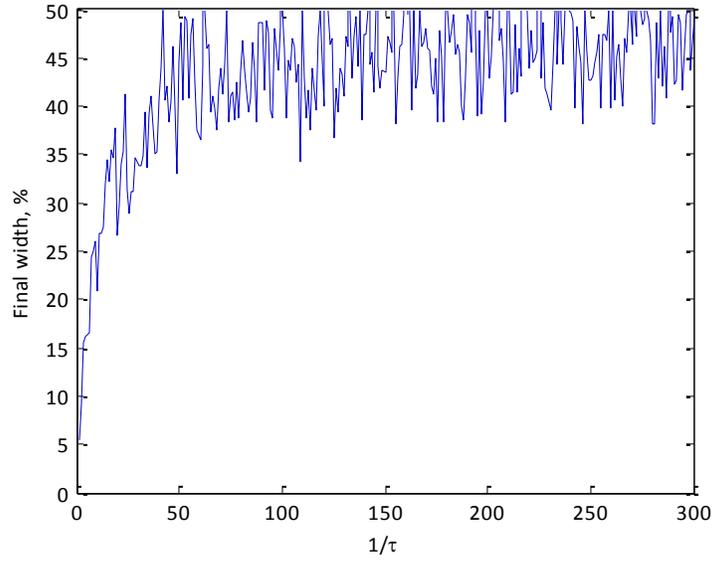

Fig. 10. Saturation of final width $w(T)$ with $\frac{1}{\tau}$.

Another way to see this is by analyzing the propagation operator

$$\hat{R} = e^{-i\hat{Q}\frac{\Delta t}{\tau}} \qquad (40)$$

Since $\hat{Q}$ is a tridiagonal matrix, $\hat{R}$ is almost tridiagonal, meaning that the only elements substantially different from zero are the diagonal and its adjacent elements [24,25]. This fact does not change if $\tau$ diminishes. However, diminishing $\tau$ randomizes the phases at the adjacent nodes. Visualization of absolute values of $\hat{R}$ is provided in Fig. 11. We can say that $R(x, x')$ are random and have standard deviation $\sim \epsilon$ in dimension $x'$. As a result, variance of the resulting distribution $|\psi(x, \Delta t)|^2$ widens by about $\epsilon$ after each step of propagation, which brings us back to Eq. (39).

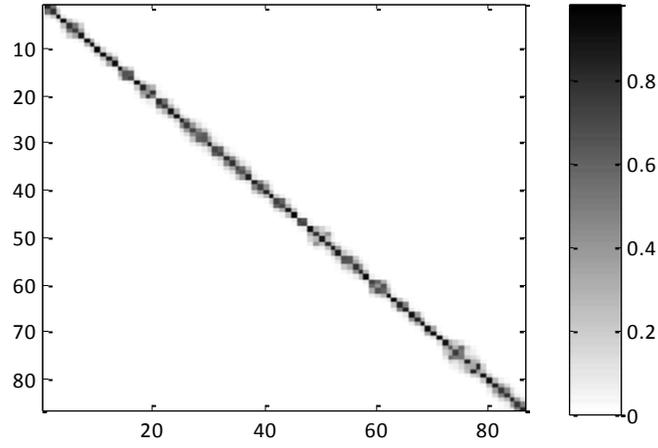

Fig. 11. Visualization of absolute values of matrix elements of $\hat{R}$.

Summarizing these considerations, we can write after $t$ steps:

$$w(t) = w_0 \sqrt{1 + \left(\frac{\beta\epsilon}{w_0}\right)^2 \frac{t}{\Delta t}} \qquad (41)$$

According to this equation, initially spread grows linearly with time. That dependence changes to $\sim\sqrt{t}$ for times longer than $t \gtrsim \left(\frac{w_0}{\beta\epsilon}\right)^2 \Delta t$. Traditional volatility can now be written through microstructural parameters as:

$$\sigma_T = \beta\epsilon \sqrt{\frac{1}{\Delta t_T}} \qquad (42)$$

Here $\Delta t_T$ is the $\Delta t$ expressed in time units of $[T]$. Practically Eq. (41) should only be applied for $t \geq \Delta t$. In order to avoid confusion, we can shape it in a more convenient form:

$$w(t) = w_{\Delta t} \sqrt{1 + \left(\frac{\beta\epsilon}{w_{\Delta t}}\right)^2 \left(\frac{t}{\Delta t} - 1\right)} \qquad (43)$$

where

$$w_{\Delta t} = w_0 \sqrt{1 + \left(\frac{\beta\epsilon}{w_0}\right)^2} \qquad (44)$$

Behavior of $w(t)$ repeats the conclusions of [2]. We see that the randomness of trading environment combined with the local character of interaction between price levels leads to diffusion of probability amplitude in price space. Localization width given by Eq. (43) allows to blend the intrinsic bid-ask spread at small time scale with the $\sqrt{t}$ diffusion-like behavior at large time scale.

## 4. Calibration to market data

In this section we demonstrate how this model works in practice. We will consider the high-low bar data as spread[4] and set the initial probability width equal to average 1-step spread. We then calibrate $\sigma_\mu$ and $\sigma_\gamma$ to fit the 1-step price spectrum. Next, we choose $\tau$ at which the spread curve $w(T)$ stabilizes. With these parameters we generate probability amplitude dynamics in a number of realizations.

Risk values (VaR and Expected Shortfall) are computed as averages over the ensemble, and not as measures of the average distribution:

$$Risk = \langle Risk^{(r)} \rangle_r \qquad (45)$$

---

[4] Possibility of a more general interpretation of spread than just plain bid-ask spread was mentioned earlier.

*Calibration to intraday 1-minute high-low data for LULU*

Figs. 12-14 present calibrated plots for minute-scale high-low bar data for Lululemon Athletica Inc. (ticker LULU) on March 16, 2016. LULU is an average liquidity NASDAQ-listed stock and on that date its 1.2 mln. shares traded with average relative bid-ask spread 0.06% and average trading time 3.2 sec.

Fig. 12a-b show a single scenario and an ensemble average of probability distribution dynamics. Fig. 13a allows to compare the initial and final probability distributions, whereas the final distributions are shown as a single scenario and ensemble average. Time evolution of probabilities of various returns is shown in Figs. 13b. Risk measures – volatility, 95% VaR, and 95% Expected shortfall – are shown in Figs. 13c. Finally, the relation of risk measures to their Gaussian benchmarks are shown in Figs. 13d. Spread curve can also be obtained directly from Eq. (43) with $w_{\Delta t} = 0.07\%$ and $\beta\epsilon = 0.001$ and is shown in Fig. 14.

We continue providing calibrated data for probability distribution of daily high-lows of LULU in Figs. 15-17. Lastly, Figs. 18-23 show calibrated data for probability distributions of 1-minute and daily high-lows for APPL.

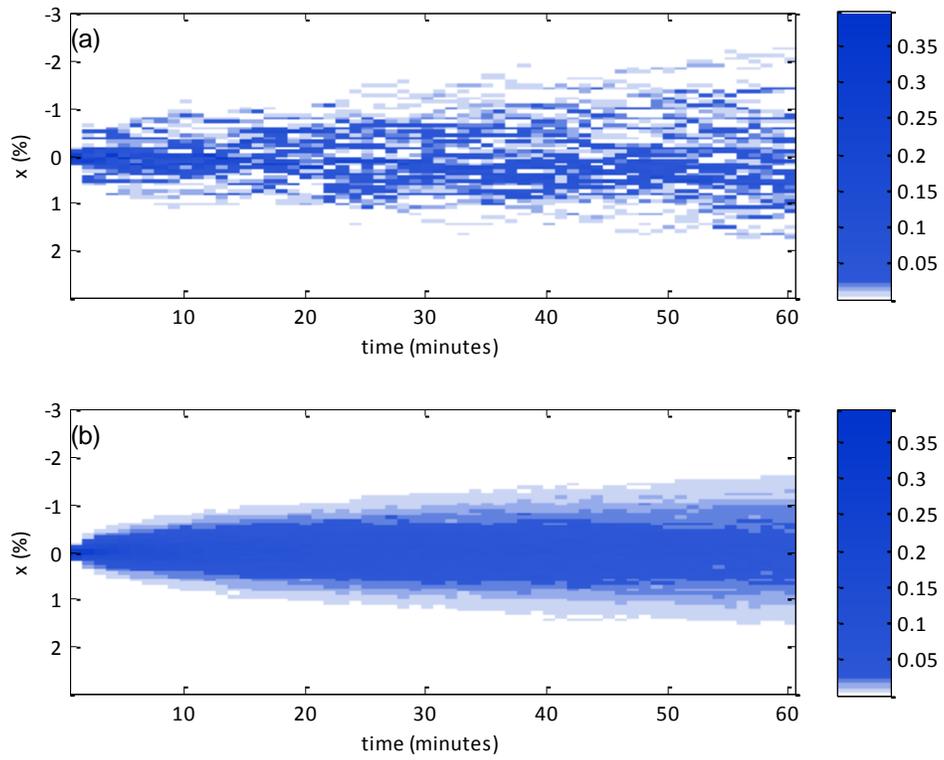

Fig. 12. Dynamics of probability distribution of intraday 1-minute high-lows for LULU ticker:

(a) single realization (b) ensemble average over 100 realizations

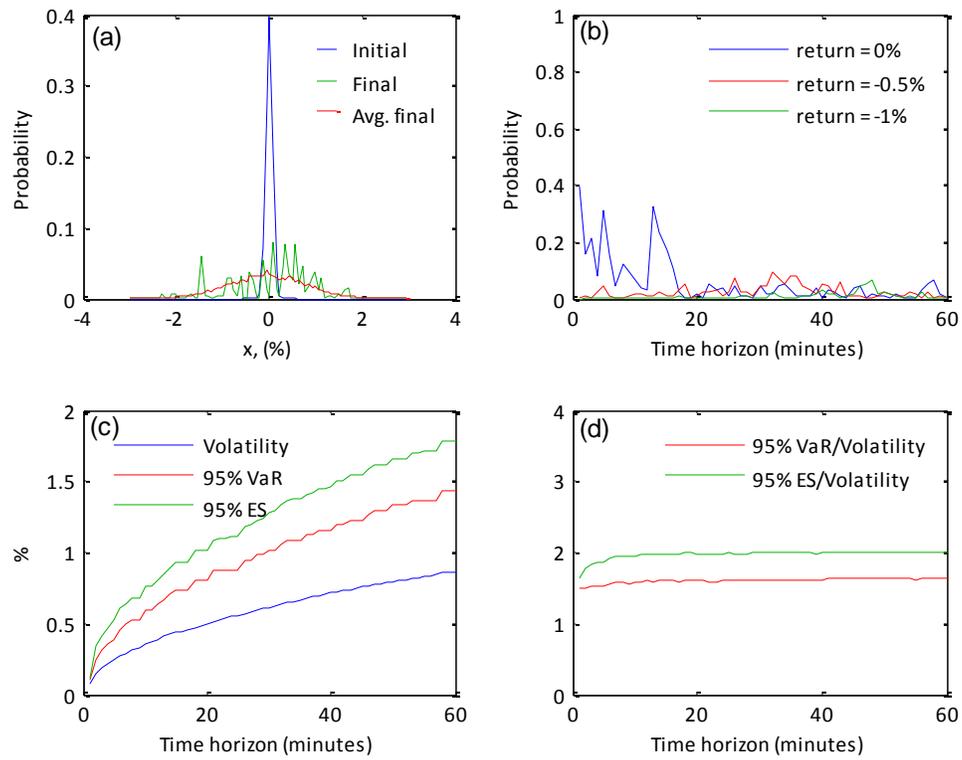

Fig. 13. Dynamics of probability distribution of intraday 1-minute high-lows for LULU ticker: (a) initial, final, and average final probability distributions, (b) dynamics of probabilities, (c) risk measures, and (d) risk ratios

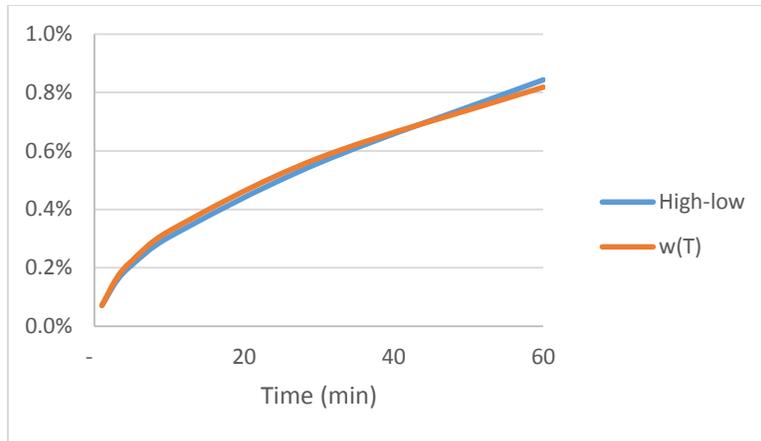

Fig. 14. Spread curve of LULU on March 16, 2016 and its fit with Eq. (43) using $w_{\Delta t} = 0.07\%$ and $\beta\epsilon = 0.001$. The average bid-ask spread was also equal 0.06%.

*Calibration to daily high-low data for LULU*

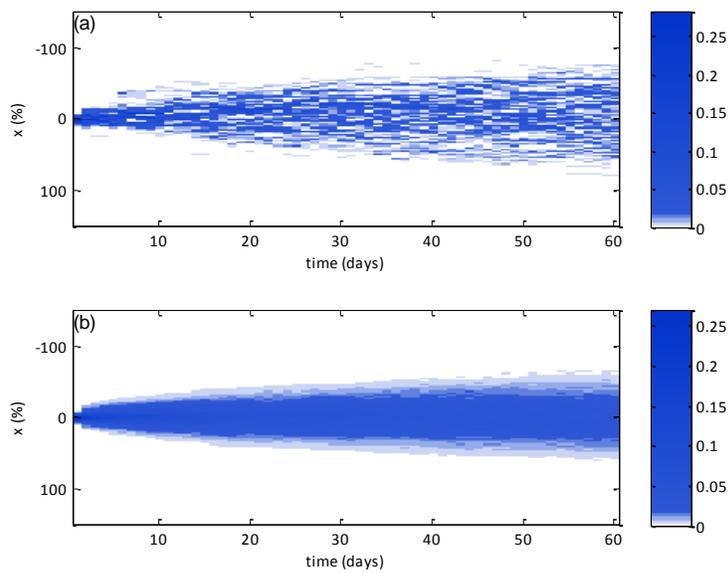

Fig. 15. Dynamics of probability distribution of daily high-lows for LULU ticker:

(a) single realization (b) ensemble average over 100 realizations

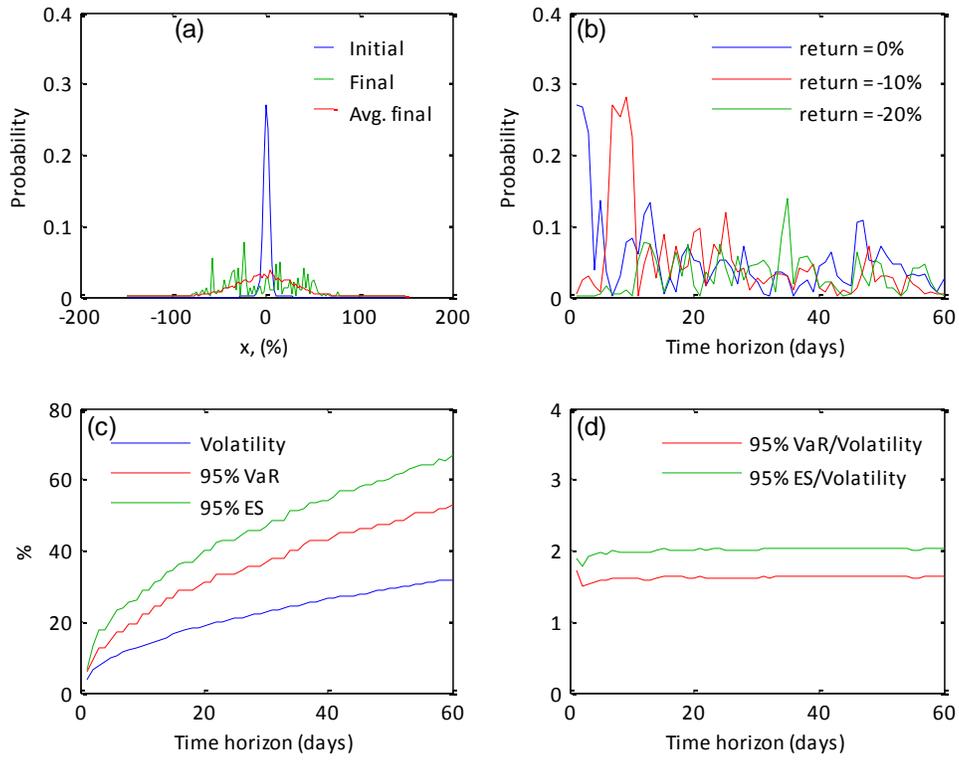

Fig. 16. Dynamics of probability distribution of daily high-lows for LULU ticker: (a) initial, final, and average final probability distributions, (b) dynamics of probabilities, (c) risk measures, and (d) risk ratios

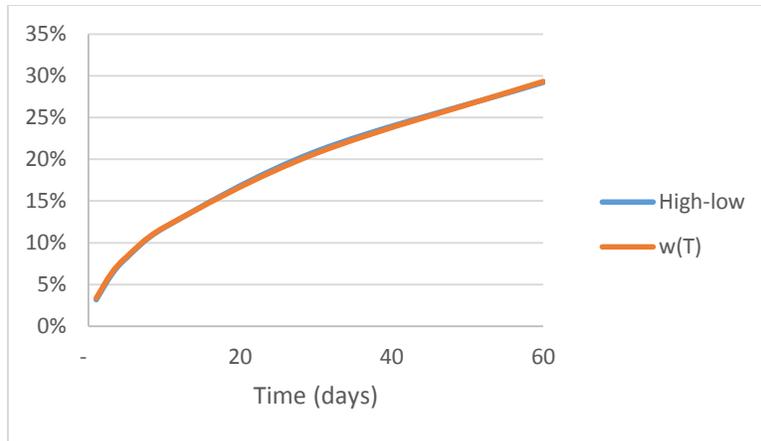

Fig. 17. Spread curve of LULU, based on daily high-low data, and its fit with Eq. (43) using $w_{\Delta t} = 3.17\%$ and $\beta\epsilon = 0.0379$. Difference is barely discernable.

*Calibration to intraday 1-minute high-low data for AAPL*

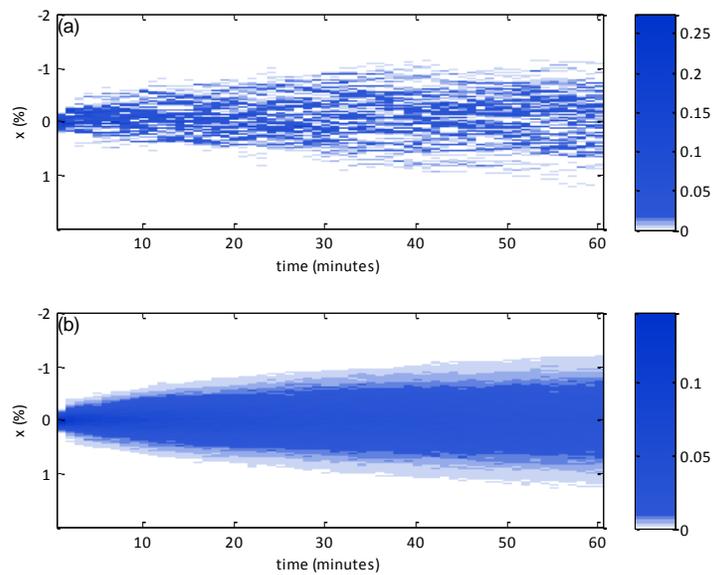

Fig. 18. Dynamics of probability distribution of intraday 1-minute high-lows for AAPL ticker: (a) single realization (b) ensemble average over 100 realizations

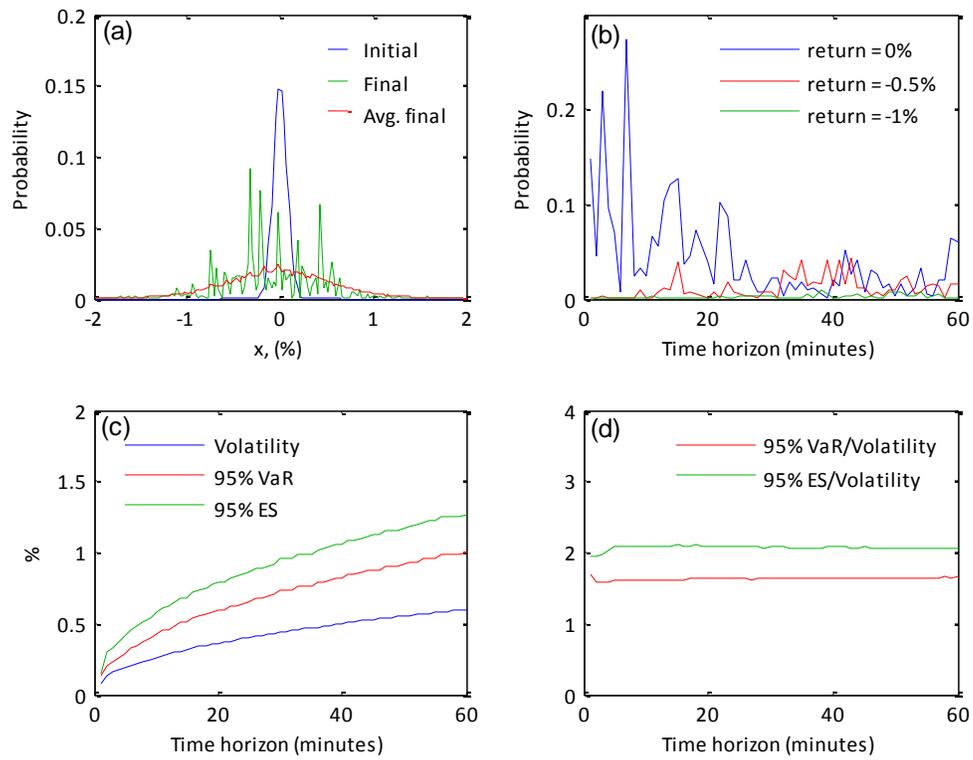

Fig. 19. Dynamics of probability distribution of intraday 1-minute high-lows for AAPL ticker: (a) initial, final, and average final probability distributions, (b) dynamics of probabilities, (c) risk measures, and (d) risk ratios

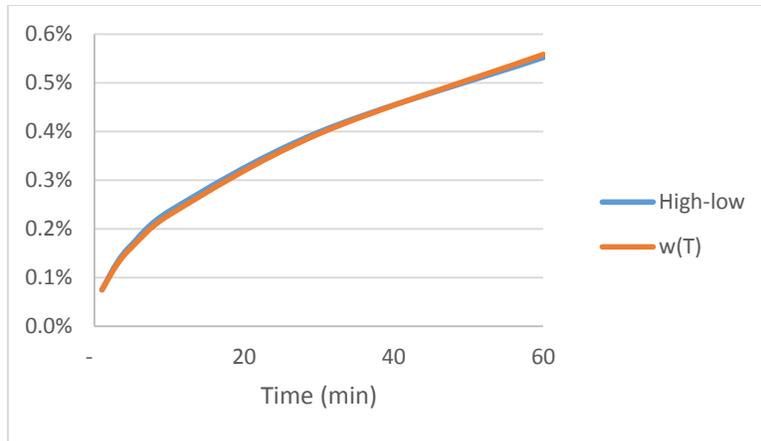

Fig. 20. Spread curve of AAPL on March 16, 2016 and its fit with Eq. (43) using $w_{\Delta t} = 0.07\%$ and $\beta\epsilon = 0.00072$. The average bid-ask spread was 0.01%, which corresponds to spread of $0.01.

*Calibration to daily high-low data for AAPL*

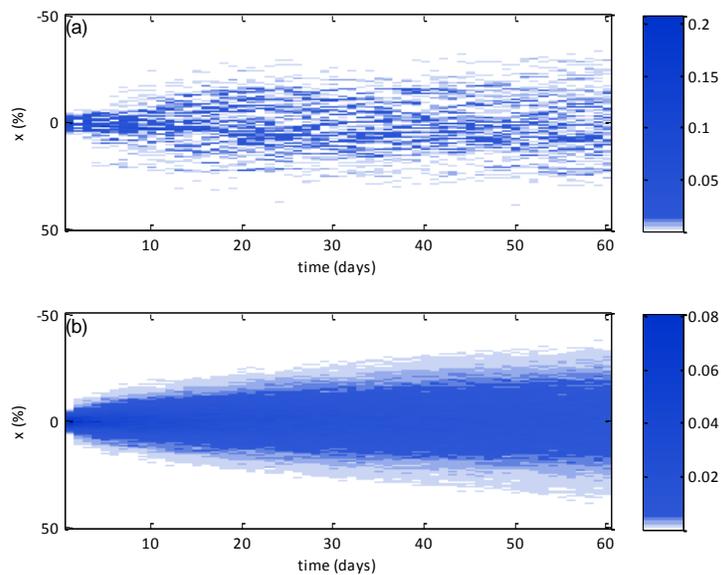

Fig. 21. Dynamics of probability distribution of daily high-lows for AAPL ticker:

(a) single realization (b) ensemble average over 50 realizations

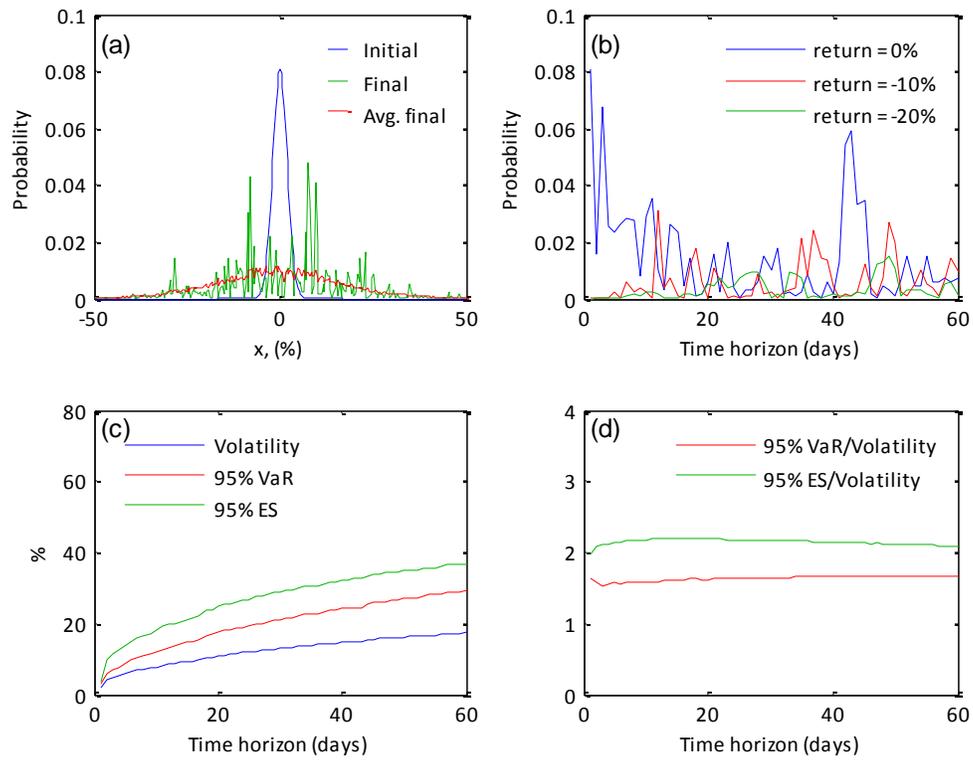

Fig. 22. Dynamics of probability distribution of daily high-lows for AAPL ticker: (a) initial, final, and average final probability distributions, (b) dynamics of probabilities, (c) risk measures, and (d) risk ratios

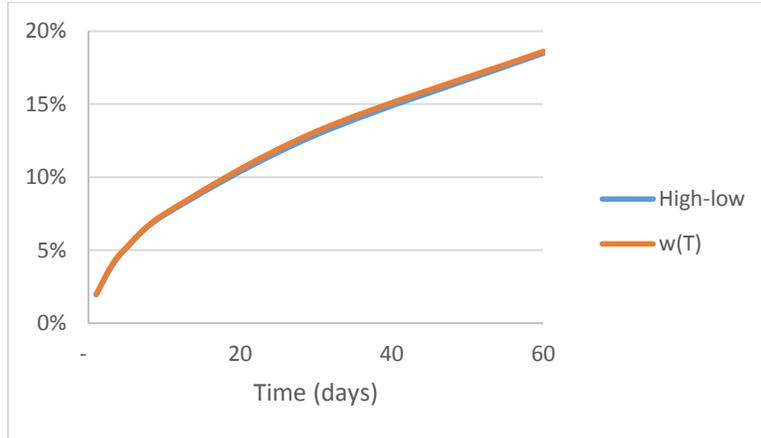

Fig. 23. Spread curve of AAPL, based on daily high-low data, and its fit with Eq. (43) using $w_{\Delta t} = 1.97\%$ and $\beta\epsilon = 0.0241$.

## 5. Discussion and conclusions

Thus, we showed that price forms as a result of selection out of the entire spectrum of prices according to probability of the security to be in each corresponding state. Price spectrum is represented by a price operator whose matrix elements fluctuate in time. In matrix form dynamics of probability distribution obeys the Schrodinger equation with properly build price operator. In log-price coordinate representation probability distribution obeys the Fokker-Planck equation with complex coefficients. Price remains localized within a range until the act of measurement, which occurs at the time of transaction.

As a result of this framework we showed that return probability distribution, usually thought to be a smooth bell shaped curve, is not actually bell shaped at all. Even though localized in price, it is erratic, and its evolution is also erratic! Only after averaging over many possible scenarios it becomes smooth. However, in reality we must face a single scenario with its erratic behavior, not an ensemble average. This in turn means that any price shift, even big, can at times build up a good chance of occurring and lead to the so called "Black Swans".

Although probability behavior is erratic the distribution width grows slowly, exhibits strong non-Gaussian features at small time scale, and the $\sqrt{t}$ behavior at large time scale. This diffusion-like behavior has its origin in randomness of phase noise introduced by the fluctuations of trading environment and the close range of price level interaction.

Despite the apparent similarity between Schrodinger and diffusion equations, one cannot be thought of as a direct extension of the other only with an imaginary diffusion coefficient[5]. Physical solutions of the two are completely different: Schrodinger equation provides solutions that are oscillatory and diffusion equation provides solutions that are dissipative. Yet, disorder of environment can alter the time dependence of the probability distribution's width and even lead to complete loss of it [26].

Up to date using quantum framework of price formation we have been able to describe the following elements of financial markets:

1) statistical behavior of spread and mid-price [1]
2) spread's scaling behavior [1,2]
3) spread curve blending the minimum spread with the $\sqrt{t}$ power law [present work]
4) relationship between spread, volatility and volume [2]
5) fat tails and rare events [present work]

Along with this progress there are difficulties as well, at least that's how they currently look. One of them is the linearity of the theory's basic equations. Linearity of equations implies validity of the superposition principle and eliminates chances of direct self-action. There is no clear evidence that this can be said of financial markets. Doubling an order size does not necessarily lead to a doubling effect on price localization. Even worse, it can start processes that were impossible with a lower order size.

---

[5] Even though analytic continuation is possible under certain constraints.

Another difficulty lies in responsive nature of financial markets. As a comparison, in measuring physical properties of elementary particles, the measuring device does not affect the quantity under measurement until it comes in contact with the object. Financial markets are different. Even in the absence of transactions traders on buy and sell sides affect each other's actions. As a result of that security price may be affected before it is measured.

An element that was seemingly left out of the theory is the order size. Order size affects price and is one of the main characteristics of a transaction. Some light has been shed on the issue in our work [2], where relationship between order size, transaction frequency, spread and volatility is studied.

It is also not obvious how quantum approach can describe situations with limited free float, or equities of companies under the acquisition process, where equity price is primarily dictated by the acquirer. We tend to think that answers to these questions will come as a result of further research.

Summarizing, we have expanded the basic framework proposed by us in [1] and developed a price formation theory based on the nature of price and price measurement process. This framework deals with spread as its intrinsic property and provides a smooth transition from small time scale when spread is important to large time scale when spread becomes unimportant.

This theory opens new capabilities for financial institutions that are involved in market-making and securities dealing activities. It allows to model price evolution in a consistent way and we have shown that behavior resulting from this framework agrees with measurable market data. This model can be calibrated to various types of data, such as best bid-and-ask, effective bid-ask, or even OHLC bars. Using the calibrated model firms and trading desks can price securities, particularly ones with limited liquidity, measure risk associated with spread, gauge it against the regular mid-price risk, and protect against rare undesired events. All these capabilities are extremely important when a trading desk's risk/return profile substantially depends on spread.